# The C($^3$P) + NH$_3$ reaction in interstellar chemistry:

# I. Investigation of the product formation channels

Jérémy Bourgalais,[1] Michael Capron,[1] Ranjith Kumar Abhinavam Kailasanathan,[2] David L. Osborn,[3] Kevin M. Hickson,[4,5] Jean-Christophe Loison,[4,5,]* Valentine Wakelam,[6,7] Fabien Goulay,[2,]* and Sébastien D. Le Picard[1,]*

[1] Institut de Physique de Rennes, UMR CNRS 6251, Université de Rennes 1,

Bât. 11C, Campus de Beaulieu, F-35042 Rennes Cedex, France

[2] Department of Chemistry, West Virginia University, Morgantown, West Virginia 26506, USA

[3] Combustion Research Facility, Mail Stop 9055, Sandia National Laboratories, Livermore, California 94551, USA

[4] Université de Bordeaux, Institut des Sciences Moléculaires, UMR 5255, F-33400 Talence, France.

[5] CNRS, Institut des Sciences Moléculaires, UMR 5255, F-33400 Talence, France.

[6] Univ. Bordeaux, LAB, UMR 5804, F-33270 Floirac, France.

[7] CNRS, LAB, UMR 5804, F-33270 Floirac, France.

*sebastien.le-picard@univ-rennes1.fr
*fabien.goulay@mail.wvu.edu
*jean-christophe.loison@u-bordeaux.fr

**ABSTRACT**

The product formation channels of ground state carbon atoms, C($^3$P), reacting with ammonia, NH$_3$, have been investigated using two complementary experiments and electronic structure calculations. Reaction products are detected in a gas flow tube experiment (330 K, 4 Torr) using tunable VUV photoionization coupled with time of flight mass spectrometry. Temporal profiles of the species formed and photoionization spectra are used to identify primary products of the C + NH$_3$ reaction. In addition, H-atom formation is monitored by VUV laser induced fluorescence from room temperature to 50 K in a supersonic gas flow generated by the Laval nozzle technique. Electronic structure calculations are performed to derive intermediates, transition states and complexes formed along the reaction coordinate. The combination of photoionization and laser induced fluorescence experiments supported by theoretical calculations indicate that in the temperature and pressure range investigated, the H + H$_2$CN production channel represents 100% of the product yield for this reaction. Kinetics measurements of the title reaction down to 50 K and the effect of the new rate constants on interstellar nitrogen hydride abundances using a model of dense interstellar clouds are reported in paper II.

*Key words*: Physical data and processes: astrochemistry, molecular processes – Interstellar medium (ISM), nebulae: atoms, molecules – Methods: laboratory: atomic, molecular

## 1. INTRODUCTION

More than 180 molecules have been identified in interstellar and circumstellar environments and many more are expected to be discovered. Although mostly observed in high mass star forming regions, a significant fraction of these molecules have been formed in dark and dense molecular clouds (mostly composed of $H_2$ and He), where the temperatures are as low as 10K. In these clouds, in addition to reactions in the gas-phase, the densities are high enough for interstellar standards (about $10^4$ cm$^{-3}$), to produce collisions between molecules in the gas-phase and interstellar grains. Those collisions and chemical reactions at the surface of the grains result in the formation of a mantle of chemical compounds. To understand the abundances of interstellar molecules and their evolution, large efforts have been made since the 1970s with the pioneering work by **Herbst and Klemperer(Herbst & Klemperer 1973)**, to propose networks of reactions that can occur in the extreme conditions prevailing in these environments. For these first chemical models, there were essentially no laboratory measurements at the temperatures of interstellar clouds (ISCs), and chemical models relied on estimates of the rate constants for any reactions that were included in the network. The last few decades have seen important developments in the fields of both experimental and theoretical kinetics to measure or predict rate constants, improving the accuracy of chemical models. At the same time the complexity of astrochemical models has increased considerably, including various physico-chemical processes (cosmic-ray particle and UV photon interactions, gas phase chemistry and grain interactions) (see Agundez and Wakelam and references therein (Agundez & Wakelam 2013)). Current reaction networks contain more than 7,000 gas-phase reactions for more than 400 atomic and molecular species, which comprise neutral, positively charged, and negatively charged species (Lee, Bettens, & Herbst 1996; Wakelam 2015; Wakelam et al. 2012; Woodall et al. 2007). However, only a small fraction of the rate constants of bimolecular reactions included in the models have been measured, especially at the low temperatures prevailing in ISCs. In addition, crucial information on the chemical identity of the products and their branching ratios is particularly scarce. In recent years, to give some guidance on the most relevant systems to study, sensitivity analyses have been used to identify elementary reactions that have the most influence on the predicted abundances of one or more observed species and their precision (Vasyunin et al. 2008; Wakelam, Herbst, & Selsis 2006; Wakelam et al. 2009; Wakelam et al. 2005).

Atomic carbon is the fourth most abundant element in the Universe and according to $C(^3P)$ observations in dense ISCs (Phillips & Huggins 1981; Schilke et al. 1995; Stark et al. 1996) it could be one of the most abundant species in these environments. Moreover, Loison et al. (2014) showed that because of the conversion reaction HNC + C → HCN + C, the abundance ratio HCN/HNC is a strong indicator of the presence of atomic carbon in dense clouds. Atomic carbon has been shown to be highly reactive down to very low temperatures (~15 K) with a variety of molecular species including small unsaturated or saturated hydrocarbons (Chastaing et al. 1999; Chastaing, Le Picard, & Sims 2000; Chastaing et al. 2001; Shannon et al. 2014). $C(^3P)$ atoms may therefore play an important role in the formation and the destruction of interstellar species.

Nitrogen is the fifth most abundant element in the Universe and is a fundamental component of molecules of biological interest. Because nitrogen atoms cannot be directly observed in the dense interstellar medium, chemical modeling and observations of N-bearing molecules have to be used to estimate their abundance. Neutral nitrogen hydrides, NH, $NH_2$ and $NH_3$ are among the first neutral nitrogen bearing molecules synthesized in dense clouds. The ratios of their abundances remain uncertain essentially because the knowledge of the chemistry of nitrogen needs improvement (Hily-Blant et al. 2010; Le Gal et al. 2014). $NH_3$ was the first interstellar molecule detected in emission by Cheung et al. in 1968 (Cheung et al. 1968). Observations of $NH_3$ in dense clouds lead to a fractional abundance **with respect to $H_2$ of ca. $10^{-8}$ (Le Gal, et al. 2014)**, at least two orders of magnitude greater than the abundances of $NH_2$ and NH.

If the reaction of ground state $C(^3P)$ atoms with $NH_3$ is efficient, it could therefore be important for determining the abundance of $NH_3$ and for a better understanding of the chemistry of nitrogen in ISCs. However, little is known on the reactivity and the products formed by the title reaction. Shevlin and co-workers were the first to study mechanisms of the $C(^3P) + NH_3$ reaction using the co-condensation of arc generated carbon vapor with ammonia on the walls of a reactor **at 77 K,** with the aim to investigate the formation of amino-acids in extraterrestrial environments (McPherson, McKee, & Shevlin 1983; Shevlin, McPherson, & Melius 1981, 1983). They deduced from their experiment that $CH_2NH$, methyleneamine, is the primary product, which may be formed by the reaction of $C(^1D)$ atoms with $NH_3$. More recently, Deeyamulla and Husain (Deeyamulla & Husain 2007) reported a kinetic study at room temperature of the reaction between $C(^3P)$ atoms and various nitrogen-containing molecules including $NH_3$, in a slow flow experimental apparatus ($T$ = 300 K, $P$ = 30 Torr). $C(^3P_J)$ atoms were generated from pulsed photolysis ($\lambda$ > 160 nm) of $C_3O_2$ in the presence of an excess of helium. $C(^3P_J)$ atoms were monitored during the reaction using the absorption at $\lambda$ = 166 nm of the ($3^3P_J$–$2^3P_J$) transition. No significant decay of the carbon atom signal due to reaction with ammonia could be observed and only an upper limit of the rate constant was derived by the authors: $k$ < $1.1 \cdot 10^{-11}$ $cm^3$ $molecule^{-1}$ $s^{-1}$. Furthermore, this experiment did not allow any information to be obtained on the products formed by the reaction.

In this article we present an investigation of the products formed by the $C(^3P) + NH_3$ reaction. Electronic structure calculations were performed to derive intermediates, transition states and complexes, formed along the reaction coordinate. Two different series of measurements were undertaken in order to unambiguously identify the preferred product channels and measure their branching fractions. Firstly, product detection experiments were performed in a flow tube reactor (330 K, 4 Torr) at the Advanced Light Source using time-resolved tunable VUV photoionization and time of flight mass spectrometry. These experiments were complemented by H-atom product yield measurements undertaken using the Laval nozzle technique coupled to laser induced fluorescence (LIF) spectroscopy down to 50 K. Kinetic experiments between 50 and 296 K as well as a study of the influence of the C + $NH_3$ reaction on nitrogen hydride abundances using a dense interstellar cloud model are reported in paper II.

## 2. EXPERIMENTAL DETAILS

*2.1 Carbon atom production*

The mass spectrometry and LIF experiments performed here do not allow us to distinguish between products formed by carbon atoms in different electronic states. It was therefore important to avoid the formation of atomic carbon in its first electronic excited state, $C(^1D)$.

As the photolysis of $C_3O_2$ at 193 nm is expected to lead only to the formation of carbon atoms in their electronic ground state, $C(^3P)$,(McFarlane et al. 1989; Okabe 1978) this method was selected for the mass spectrometry experiments. Carbon suboxide was synthesized via the procedure reported by (Capron et al. 2015).

Side reactions of the photolysis products of carbon suboxide, such as $C_2O(\tilde{a}\ ^1\Delta)$, are likely to generate a significant amount of H atoms through secondary reactions. For this reason, $CBr_4$ was chosen as the source for carbon atoms for the H-atom LIF detection experiments. Previous work (Shannon, et al. 2014) has shown that excited state $C(^1D)$ atoms are also produced by the multiphoton dissociation of $CBr_4$ (10-15% with respect to $C(^3P)$). To avoid potential H atom formation from the $C(^1D)$ + $NH_3$ reaction, H atom yield measurements were either performed using pure nitrogen based flows (296 K, 177 K and 106 K), or using an argon based flow (50 K) with a large concentration of added $N_2$ (1.6 · $10^{16}$ molecule $cm^{-3}$). $N_2$ is known to quench $C(^1D)$ atoms efficiently (Husain & Kirsch 1971) thereby preventing the simultaneous production of H atoms from $C(^1D)$ reactions. For room temperature experiments performed in 5 Torr of $N_2$, the characteristic decay time (half-life) for relaxation of $C(^1D)$ was approximately 1 μs.

*2.2 Tunable VUV photoionization mass spectrometry*

The reaction is performed in a gas mixture flowing slowly through a 62 cm quartz tube sampled with a time-resolved photoionization time of flight mass spectrometer at the Advanced Light Source synchrotron. A description of the apparatus has been given elsewhere (Osborn et al. 2008; Taatjes et al. 2008) and only a brief overview is presented here. The helium buffer gas (99.999 %), precursor, $C_3O_2$, and reaction gas, $NH_3$ (Alpha Gaz, N36), are introduced in the reaction flow via separate mass-flow controllers. The pressure in the reactor was typically 4 Torr with a total gas flow of 250 sccm. All the experiments are performed at 333 K with a total flow density of 1.16 · $10^{17}$ $cm^{-3}$. $C(^3P)$ reactions are initiated by the photolysis of carbon suboxide, $C_3O_2$, at 193 nm using a pulsed excimer (ArF) laser operating at 10 Hz, allowing sufficient time to completely refresh the gas mixture between laser pulses. The laser power output is 266 mJ per pulse (20 ns), corresponding to a fluence inside the reaction flow tube of ~ 20 to 50 mJ/$cm^2$. The concentrations of $C_3O_2$ and ammonia introduced were typically $10^{14}$ and $10^{13}$ $cm^{-3}$ respectively and we estimate the concentration of carbon atoms to be $10^{11}$ $cm^{-3}$ (Capron, et al. 2015).

A 650 $\mu$m pinhole in the side of the tube, ~30 cm along the flow, is used to sample the gas. The effusive beam emerging from this pinhole is skimmed before entering an ionization chamber where it intersects the quasi-continuous vacuum-ultraviolet (VUV) synchrotron radiation. Ions formed are detected as a function of their mass-to-charge (m/z) ratio using a 50 kHz pulsed orthogonal-acceleration time-of-flight mass spectrometer. Complete mass spectra are then collected as a function of reaction time. The total time-window is set to be 50 ms with ion detection 10 ms before and 40 ms after the laser pulse.

*2.3 H atom detection using VUV LIF*

The H atom product formation experiments were conducted with a small continuous supersonic flow reactor adapted from the original apparatus designed by (Rowe et al. 1984). A description of this technique has been reported in earlier papers (see for example (Daugey et al. 2008)). Three Laval nozzles were used in this investigation allowing flow temperatures of 177 K, 106 K and 50 K to be achieved. These values are given in Table 1.

Table 1 Characteristics of the supersonic flows

| Mach number | $1.8 \pm 0.02^{(a)}$ | $3.0 \pm 0.02$ | $3.9 \pm 0.1$ |
| --- | --- | --- | --- |
| Carrier gas | $N_2$ | $N_2$ | Ar |
| Density ($\cdot 10^{16}$ cm$^{-3}$) | 9.4 | 10.3 | 25.9 |
| Impact pressure (Torr) | 8.2 | 13.4 | 29.6 |
| Stagnation pressure (Torr) | 10.3 | 39.7 | 113 |
| Temperature (K) | $177 \pm 2$ | $106 \pm 1$ | $50 \pm 1$ |
| Mean flow velocity (ms$^{-1}$) | $496 \pm 4$ | $626 \pm 2$ | $505 \pm 1$ |

[a] The errors on the temperature, Mach number and mean flow velocity (one standard deviation) are calculated by measurements of the impact pressure as a function of distance from the Laval nozzle and the stagnation pressure.

C($^3$P) atoms were produced by the dissociation of precursor $CBr_4$ molecules for the reasons outlined in section 2.1, through a multiphoton process using approximately 20 mJ of 266 nm radiation at 10 Hz.

The photolysis laser was steered along the reactor axis, creating carbon atoms within the supersonic flow. To carry the $CBr_4$ molecules into the reactor, a small $N_2$ or Ar flow was passed over solid $CBr_4$ upstream of the nozzle reservoir. We calculate that the maximum $CBr_4$ concentration used in these experiments was $4 \cdot 10^{12}$ molecule cm$^{-3}$, estimated from its saturated vapor pressure at room temperature.

Product H($^2$S) atoms were detected by VUV LIF through the 1s $^2$S → 2p $^2$P$^0$ Lyman-α transition at 121.567 nm and a VUV PMT. Narrowband radiation at 729.4 nm produced by a pulsed dye laser was frequency doubled, producing ultraviolet (UV) radiation at 364.7 nm. This beam was focused into a cell containing 210 Torr of krypton with 550 Torr of argon added for phase matching positioned perpendicularly to the cold flow and the PMT. The tunable VUV light produced by third harmonic generation was collimated before crossing the supersonic flow.

To determine absolute H atom yields over the 50-296 K temperature range, the H atom VUV LIF signal intensity from the C($^3$P) + $NH_3$ reaction was calibrated using the C($^3$P) + $C_2H_4$ reaction whose H-atom yield has been measured to be equal to 0.92 ± 0.04 at 300 K in a fast flow reactor experiment (Bergeat & Loison 2001). The pressure dependence of the H atom yield of the C($^3$P) + $NH_3$ reaction at 296 K was also examined in the present study. VUV LIF signals were recorded as a function of delay time between the photolysis laser and the probe laser. To ensure that diffusional losses did not lead to large errors in the estimation of the H atom yields, only profiles with similar time constants were used to extract relative intensities. As a result, excess reagent concentrations ($NH_3$ or $C_2H_4$) were carefully chosen to obtain comparable first-order-production rates for both reactions. For a given coreagent concentration, 30 datapoints were acquired at each time step with at least 50 time intervals for each formation curves. In addition, several points were recorded at negative time delays to fix the baseline level.

None of the gases used were purified before use, being flowed directly from cylinders. Digital mass flow controllers were used to control the carrier gas flows. These controllers were calibrated prior to usage for the specific gas used.

*2.3 Electronic Structure Calculations*

The potential energy surface (PES) of the $C + NH_3$ reaction has been studied at different levels of theory. Firstly, using Davidson corrected multi-reference configuration interaction (MRCI + Q) with complete active space self-consistent field (CASSCF) wave-functions associated to the aug-cc-pVQZ basis. Secondly, using DFT calculations with the M06-2X functional (Zhao & Truhlar 2008) associated to the cc-pVQZ basis. Calculations were also performed at the CCSD(T)/aug-cc-pVQZ level. The CCSD(T) and MRCI+Q calculations were performed using the MOLPRO 2010 package. DFT calculations were carried out with the Gaussian09 package. The CASSCF and MRCI calculations were performed with 13 electrons distributed in 10 orbitals with the 1s orbitals of carbon and nitrogen kept doubly occupied but fully optimized. The CASCCF calculations lead to mono-configurational wave-functions for adducts and transition states. CCSD(T) calculations should be then accurate as also indicated by the T1 and D1 diagnostic values (T1 values are in the 0.011 - 0.017 range and D1 values are in the 0.019 - 0.039 range except for TS2 with T1 = 0.042 and D1 = 0.127 leading to larger uncertainties). The geometries for the various stationary points were optimized for each method. The frequencies were calculated at the DFT and CCSD(T) levels only. We have verified that each TS was characterized by one imaginary frequency. We also determined for each TS the minimum energy pathways using intrinsic reaction coordinate analyses at the DFT level. The schematic energy diagram is shown in Figure 1. The calculated stationary point energies, geometries and frequencies of the C⋯NH$_3$ complex, TS1 and TS2 are available as online only information.

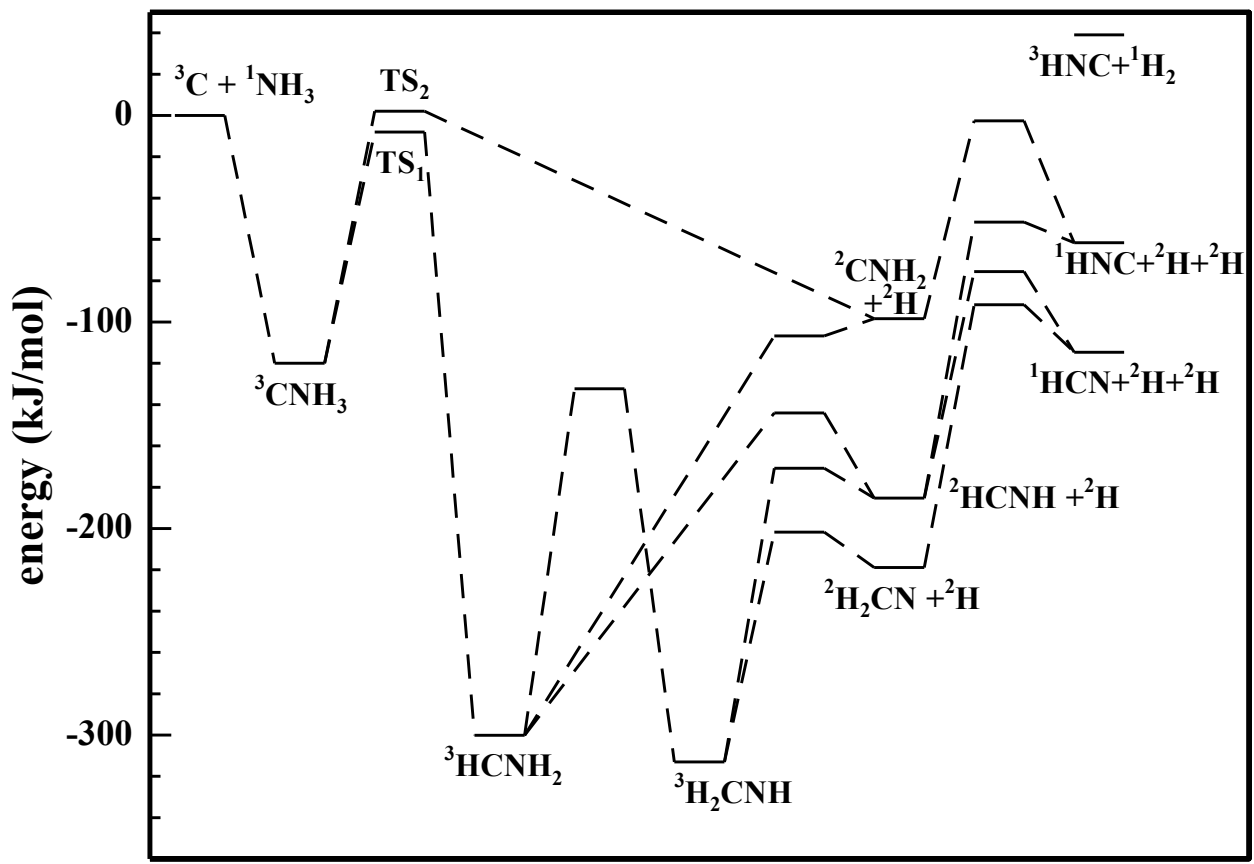

**Figure 1.** Potential energy diagram for the C($^3$P) + NH$_3$ reaction on the triplet surface. See online-only information for the relative energies.

### 3. RESULTS

*3.1 Electronic Structure Calculations*

The C($^3$P$_{0,1,2}$) + NH$_3$($^1$A$_1$) system correlates with 3 surfaces, one $^3$A′ and two $^3$A′′ surfaces in C$_s$ symmetry. At the MRCI+Q/aug-cc-pVQZ level with CASSCF geometry optimized, only the first surface is attractive, leading to a van der Waals complex. This surface is also attractive at the CCSD(T)/aug-cc-pVQZ level. The C⋯NH$_3$ complex, with the carbon atom attached to the nitrogen atom, is strongly bound (111 kJ/mol at CCSD(T)/aug-cc-pVQZ level) with a C-N bond length between 1.60 Å and 1.66 Å depending on the method used. Further evolution of the C⋯$NH_3$ complex leads to an $^3HCNH_2$ adduct through TS1 corresponding to C insertion into a N-H bond, or directly to $^2CNH_2$ + $^2H$ products through TS2.

It should be noted that the calculated complex and TS energies are very method dependent as shown in Table S1. Indeed DFT methods usually underestimate energy barriers (Wagner 2002), even if the M06-2X functional generally leads to better agreement with experimental data than the B3LYP functional. On the other hand, MRCI methods tend to overestimate energy barriers. In this study the CCSD(T) method should lead to the more accurate values considering the T1 and D1 diagnostic values (Lee 2003; Leininger et al. 2000). The uncertainties on the calculated energies are difficult to estimate precisely. The uncertainty may be around 10 kJ mol$^{-1}$ for the C⋯$NH_3$ adduct and can be notably higher for transition states. Then the importance of the pathway through TS2 is difficult to estimate as the calculated energy is close to the reagent energy at the CCSD(T) level.

*3.2 Synchrotron Experiments*

Time- and energy-resolved mass spectra are obtained by averaging at least 500 laser shots at each photon energy. In the present study the ionizing photon energy is scanned from 9.4 to 10.5 eV in steps of 0.025 eV. Ar is used as a gas filter to absorb higher energy photons. All data are normalized to account for variations of the ALS photon flux using a NIST-calibrated photodiode (SXUV-100, International Radiation Detectors, Inc.). Photoionization spectra at a given m/z ratio are obtained by integrating the three-dimensional data set over the desired mass and appropriate time window. Three independent datasets are recorded and averaged. The error bars for a given data point are twice the standard deviation around the mean value. The photon energy resolution is determined to be of the order of 40 meV by measuring the atomic resonance of Xe.

The photolysis of carbon suboxide, $C_3O_2$, at 193 nm was used in the present experiments to generate ground state carbon atoms, $C(^3P)$. The relative yield of carbon atoms to $C_2O$ following the photolysis of $C_3O_2$ at 193 nm was measured in 1989 by McFarlane et al.(McFarlane, et al. 1989) : the one-photon photodissociation of $C_3O_2$ to $C_2O$ + CO and C + 2CO was shown to be the dominant pathways, with a ratio of ground state carbon atoms, $C(^3P)$, to $C_2O$ of 0.06 ± 0.03. The absence of any significant amount of $C(^1D)$ state in our experimental conditions was demonstrated in a previous work (Capron, et al. 2015). It should be noted that the spin allowed exit channel of $C_3O_2$ photodissociation leads to $CO(X^1\Sigma^+) + C_2O(\tilde{a}\ ^1\Delta)$, an excited electronic state of $C_2O$ (Anderson & Rosenfeld 1991).

The mass spectra and photo-ionization spectra presented below are obtained upon irradiation of a mixture of carbon suboxide with ammonia. The time behaviour of the ion signals after the laser pulse is used to discriminate between $C(^3P)$ reaction products and products from side reactions.

A mass spectrum upon irradiation at 193 nm of $NH_3$ in helium recorded at 13.1 eV (not shown here) indicated the presence of $NH_3$ and $NH_2$ only. In a previous study (Capron, et al. 2015), a mass spectrum was recorded at 11.3 eV in the presence of $C_3O_2$ in helium, showing the photolytic production of $C_2O$ at m/z = 40 as the main product. Figure 2 displays an image of ion signal intensities as a function of reaction time and mass at 13.7 eV photon energy following 193-nm photolysis of a mixture of $C_3O_2$ and ammonia in helium. The image is not baseline subtracted for the signal before the laser pulse. This image allows us to determine the species appearing before or after the laser shot as well as their kinetic behaviour. The main ion signals appearing after the laser shot are at m/z = 16, 27, 28, 29 and 40. The very fast decay of signal at m/z = 40 (less than 2 ms) suggests a reactive species, likely $^1C_2O$, generated by the 193-nm photolysis of $C_3O_2$. The signal at m/z = 28 does not originate from CO molecules which are also formed by $C_3O_2$ photolysis as their ionization energy is equal to 14.01 eV. Note the absence of any signal at m/z = 12, corresponding to carbon atoms, due to both the small amount formed by the photolysis of $C_3O_2$ and their very high reactivity. Mass spectra recorded at various energies (not shown here) allowed us to identify the ion signal at m/z = 16 as originating from $NH_2$ (IE = 11.17eV (Dixon, Feller, & Peterson 2001)). $NH_2$ is known to be a product of $NH_3$ photolysis at 193 nm (Leach, Jochims, & Baumgärtel 2005). From the photoionization spectra we estimate its concentration to be less than 10% of that of ammonia, assuming the same ionization cross sections for both molecules.

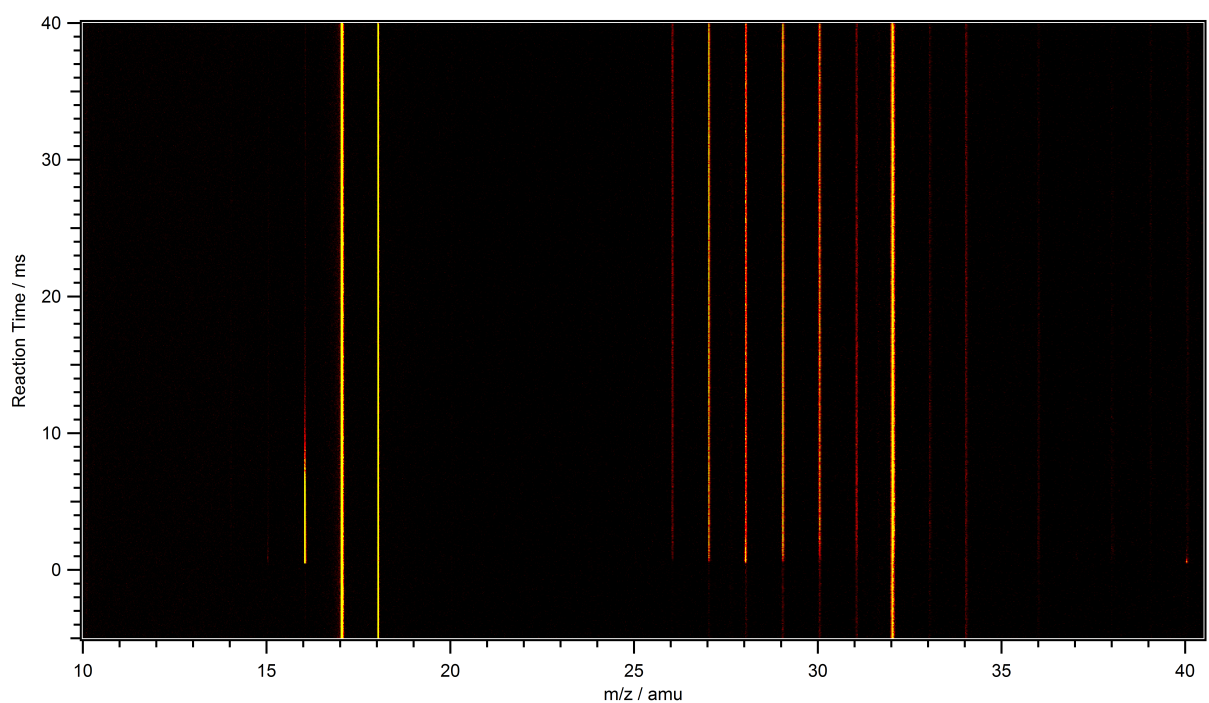

**Figure 2.** 3D mass spectrum obtained by photolysis of a carbon suboxide and ammonia mixture in helium at 13.7eV from -5 to 40 ms relatively to the laser pulse.

Figure 3 shows the non-normalized temporal traces of signals at m/z = 16, 27, 28, 29 and 40 recorded at 13.7 eV over the 0-2 ms reaction time window. Each point is the average of 2,500 acquisitions. Within the experimental time resolution, the signal at m/z = 28 corresponds to the first product formed after irradiation of the gas mixture along with the photoproducts $NH_2$ (m/z = 16) and $C_2O$ (m/z = 40). Based on their slower rise, ion signals at m/z = 27 and 29 are likely to originate from slower or secondary reactions.

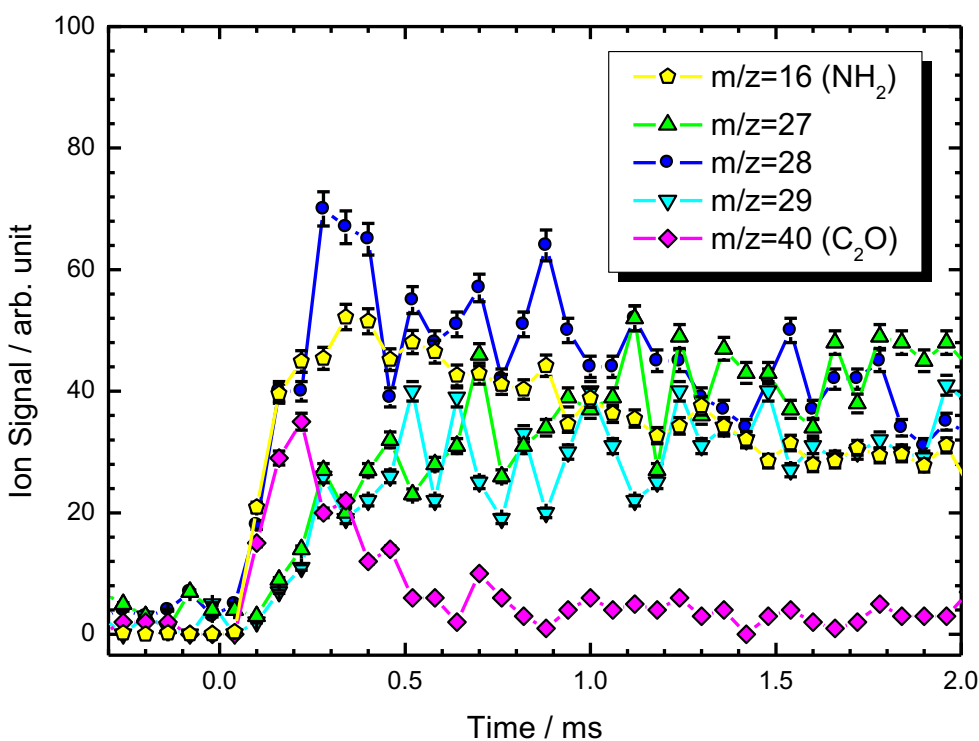

**Figure 3.** Non-normalized kinetic traces of ion signals at m/z =16 (yellow pentagons), 27 (green triangles), 28 (blue circles), 29 (cyan triangles) and 40 (pink diamonds) at 13.7 eV.

The measured photoionization spectrum for the m/z = 28 is shown in Figure 4. The solid red line is the photoionization spectrum of the $H_2CN$ radical obtained by Nesbitt et al. (Nesbitt et al.

1991a). Considering the signal to noise ratio, the reasonably good match between the two photoionization spectra indicates that $H_2CN$ is likely to be the main isomer detected at m/z = 28. The cis and trans HCNH isomers, which could also be formed by the title reaction, have calculated IEs of 6.8 and 7.0 eV respectively (Nesbitt et al. 1991a). Nevertheless, the non-observation of signal at or around these values is a strong indication that these species are not produced by the $C(^3P) + NH_3$ reaction (cf. Figure 4).

The production of $H_2CN$ by the reaction $C(^3P) + NH_3$ is consistent with the electronic structure calculations, in which the lowest energy pathway leads to formation of $H_2CN$ through TS1 after $C\cdots NH_3$ complex formation.

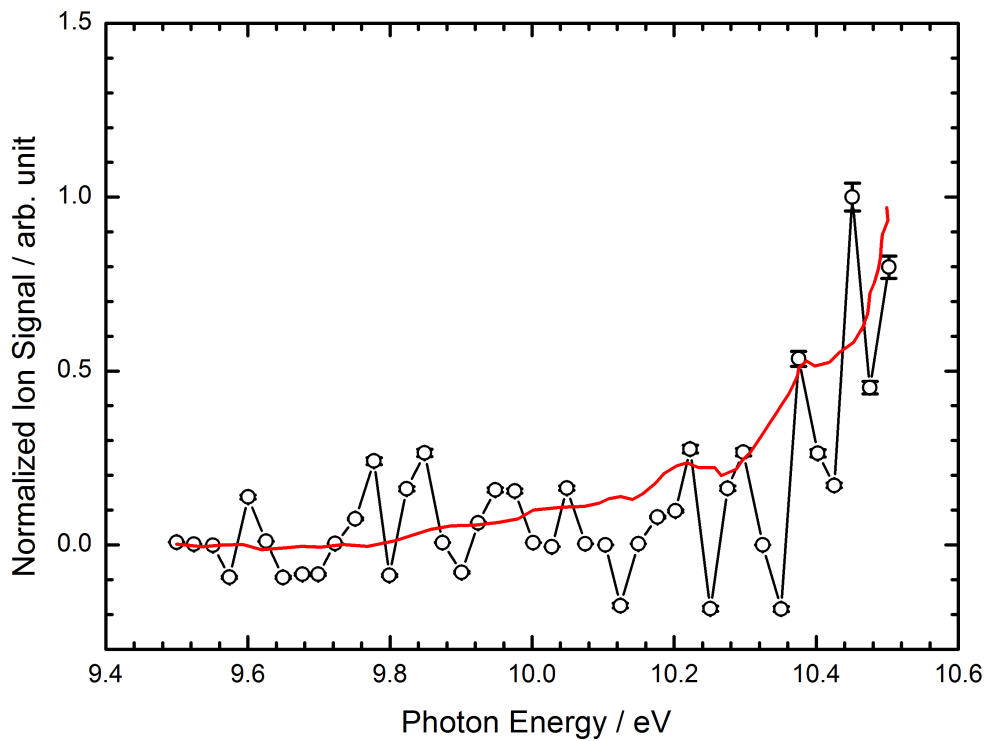

**Figure 4.** Experimental photoion spectrum integrated over 0-3 ms at m/z = 28 obtained in this work (black circles), compared with the photoion spectrum of H$_2$CN radical obtained by Nesbitt et al. (solid line) (Nesbitt, et al. 1991a).

In order to understand the formation kinetics of masses 27, 28 and 29, a reaction network was used to simulate the chemistry occurring in the reactor upon irradiation of the C$_3$O$_2$/NH$_3$ mixture. As C, C$_2$O, NH$_2$ and H are the main radicals formed by the photolysis of the gas mixture, they are considered to initiate the chemistry in the flow tube, reacting with C$_3$O$_2$ and NH$_3$. Appendix A presents the reactions included in the chemical network along with their rate coefficients taken from the literature when available, or estimated. The reactivity of the $^1$C$_2$O radical is not well known but it is likely to be high considering the very fast decay of the observed peak at m/z = 40 as shown in Figure 3. M06-2X/cc-pVTZ calculations show that $^1$C$_2$O reacts without a barrier with NH$_3$ (see potential energy diagram in Figure 5). Furthermore we assume reasonably that $^1$C$_2$O reacts without a barrier with unsaturated C$_3$O$_2$. The chemical network suggests that HCN/HNC can be produced by various secondary reactions, which can explain the kinetics of formation of the m/z = 27 ion signal. In addition to the reaction of C($^3$P) with NH$_3$, H$_2$CN (m/z = 28) could be formed by the reaction of $^1$C$_2$O with NH$_2$. Due to the efficient reactions of $^1$C$_2$O with the most abundant species (NH$_3$, C$_3$O$_2$) in the flow and the slow decrease of NH$_2$ signal intensity observed experimentally (see m/z = 16 in Figure 3) the concentration of $^1$C$_2$O can be assumed to be small relative to that of NH$_2$ and $^1$C$_2$O + NH$_2$ can be considered to be pseudo first-order. An estimation of the rate coefficient (see appendix A) leads to a characteristic time of at least 5 ms for this reaction, eliminating its contribution to the fast signal increase (less than 1 ms) observed at m/z = 28 in Figure 3. Furthermore, the reaction of C$_2$O with NH$_2$ almost certainly leads to the formation of the H$_2$NC isomer (m/z = 28) whose isomerisation towards H$_2$CN presents a barrier (Sumathi &

Nguyen 1998): the $C_2O$ + $NH_2$ reaction leads very likely to the O=C=C-$NH_2$ adduct, which should quickly evolve toward CO + $CNH_2$. Finally, the ion signal at m/z = 29 could be explained by the formation of $CH_2NH$ via the reactions of $C_2O$ with $NH_3$ (see Figure 5) or CH with $NH_3$.

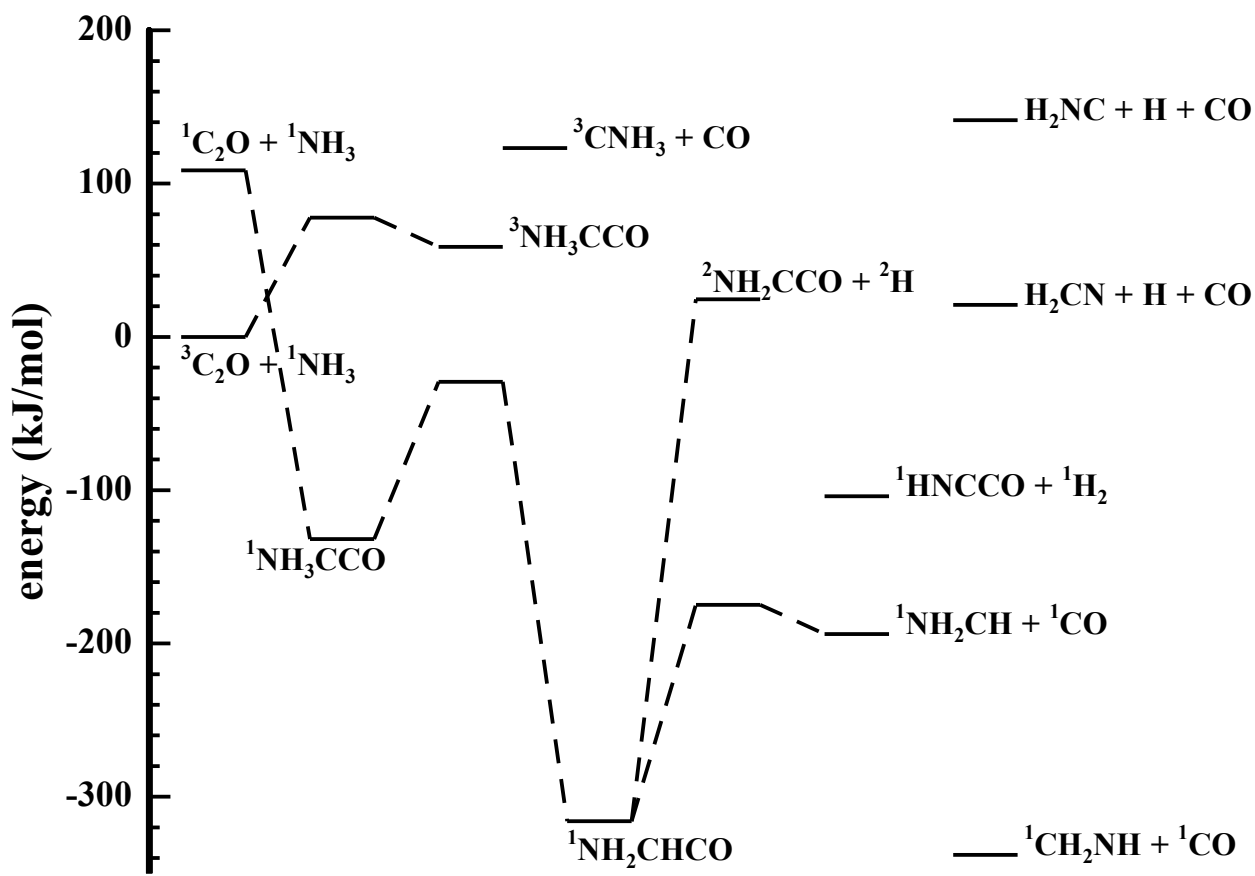

**Figure 5.** Potential energy diagram for the $C_2O$ + $NH_3$ reaction on the singlet and triplet surfaces calculated at the M06-2X/cc-pVTZ level.

The photoionization spectrum of m/z = 29 measured in this work (not shown here), is in excellent agreement with the photoionization spectrum previously measured by Goulay and co-workers (private communication) when studying the reaction between CH and $NH_3$ with the same experimental apparatus. According to its ionization energy (IE= 9.88 ± 0.07 eV (Tarasenko et al. 1986)) this product is likely to be methanimine ($CH_2NH$). It should be noted that another source of the m/z = 28 signal at time delays beyond 5 ms was observed and can be attributed to the dissociation of the second ionic state of $CH_2NH$ (IE = 12.5 eV, see Appendix C) to form $HCNH^+$ + H. An estimation of the ionization energy of the C$\cdots NH_3$ adduct (First IE = 7.7 eV and second IE = 9.4 eV using Electron Propagator Theory (EPT), with Gaussian09), allows us to rule out its presence in our experimental conditions.

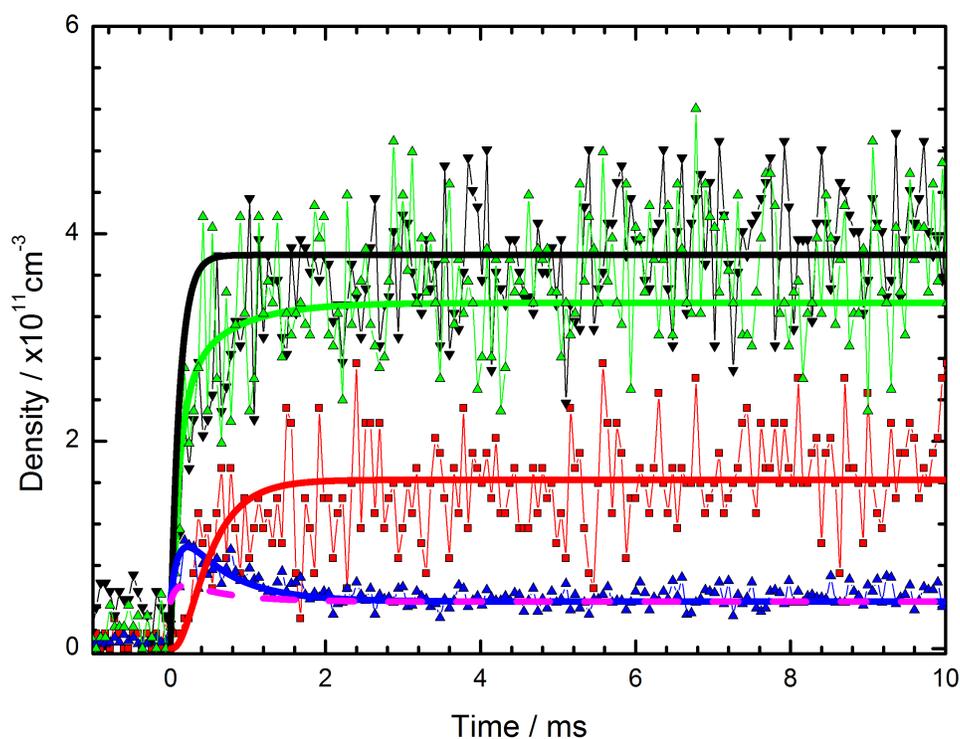

**Figure 6.** Kinetic traces normalized by ionization cross section of ion signals at m/z = 26 (red squares), 27 (green triangles), 28 (blue circles), 29 (black triangles) recorded at 13.7 eV. The solid lines are the kinetic model results based on the chemical network reported in Appendix A. The pink line shows the temporal trace of $H_2CN$ resulting from the kinetic model when removing the $C(^3P) + NH_3$ reaction.

Figure 6 displays the kinetic traces of m/z = 26, 27, 28 and 29 for a high $NH_3$ concentration (where the $H_2CN$ contribution from the C + $NH_3$ reaction is expected to be the greatest) together with the results from the kinetic model based on the chemical network presented in Appendix A. The orange dashed line represents the modelled m/z = 28 temporal trace without the contribution of the $C(^3P)$ + $NH_3$ reaction, showing that this reaction is a major source of the signal at m/z = 28 at short times. Note that for low $NH_3$ concentrations, unidentified secondary sources of the signal at m/z = 28 led to poor agreement with the simulations using the present chemical network.

In order to confirm $H_2CN$ + H as the main exit channel of the $C(^3P)$ + $NH_3$ reaction, suggested by this kinetic analysis of the photoionization mass spectrometry data and the electronic structure calculations, we performed another series of experiments using a CRESU apparatus in which H atom production is monitored by VUV LIF, described in the following section.

*3.3 CRESU Experiments*

Examples of the $H(^2S)$ VUV LIF intensity profiles recorded sequentially for the C + $NH_3$ and C + $C_2H_4$ reactions at 106 K are shown in Figure 7.

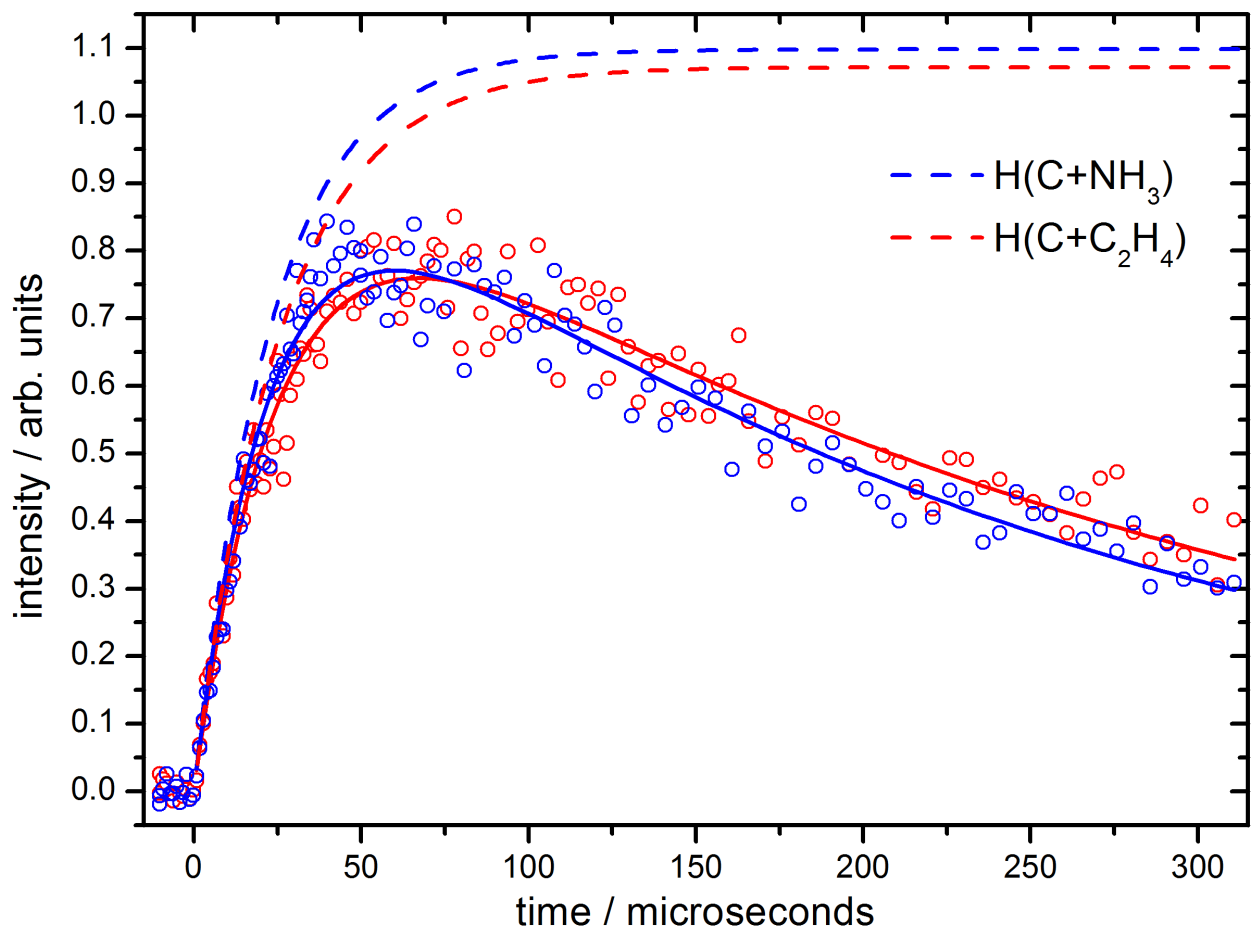

**Figure 7.** H($^2$S) VUV LIF emission profiles at 106 K. (Blue open circles) H atom signal from the C + NH$_3$ reaction with [NH$_3$] = 1.5 · 10$^{14}$ molecule cm$^{-3}$; (blue solid line) biexponential fit to the C + NH$_3$ reaction H atom data; (blue dashed line) theoretical H atom yield from the C + NH$_3$ reaction in the absence of competing H atom losses. (Red open circles) H atom signal from the C + C$_2$H$_4$ reaction with [C$_2$H$_4$] = 5.4 · 10$^{13}$ molecule cm$^{-3}$; (red solid line) biexponential fit to the C + C$_2$H$_4$ reaction H atom data; (red dashed line) theoretical H atom yield from the C + C$_2$H$_4$ reaction in the absence of competing H atom losses.

Several pairs of H-atom curves were acquired at each temperature to reduce the measurement uncertainty. In addition, the acquisition order was also varied to reduce errors arising from possible changes in the fluorescence intensities over time. These traces have two component parts; a rapid initial increase, due to the formation of hydrogen atoms followed by a slowly decaying part due to atomic hydrogen loss from the probe volume. For these experiments, a biexponential function of the form

$$I_H = A\{\exp(-k_{L(H)}t) - \exp(-k_{1st}t)\} \tag{1}$$

was used to describe the H atom signal intensity ($I_H$) as a function of time with the exponential loss term $k_{L(H)}$ to describe secondary H atom loss (through diffusion and other secondary loss processes). Here, the apparent first-order formation rate $k_{1st} = k_{C+X}[X] + k_{F(H)} + k_{L(C)}$ where X represents either $NH_3$ or $C_2H_4$ and $k_{F(H)}$ describes any additional H atom formation processes. The H atom formation rate also depends on the secondary loss of atomic carbon $k_{L(C)}$. It is this final term that constrains us to use similar first-order-production rates for both reactions, because some carbon atoms could be lost without reacting. The parameter $A$ represents the H atom VUV signal intensity in the absence of secondary H atom losses. In Figure 7, we also plot the dependence of the H atom signals we obtain when the value of the $k_{L(H)}$ term is set to zero (ie the theoretical H atom yield). The ratio of the $A$ parameter values derived from fits to the H atom signals thus represents the relative H atom yields for the two reactions. Two factors needed to be considered to obtain the absolute H atom yield for the $C + NH_3$ reaction from the relative one. Firstly, as mentioned in the experimental section, the $C + C_2H_4$ reaction has been measured to result in a branching fraction of $0.92 \pm 0.04$ for the $C_3H_3 + H$ product channel at room temperature (Bergeat & Loison 2001). This reaction is not expected to display marked temperature and pressure dependences given the absence of an initial van der Waals complex and the presence of low submerged barriers leading from the $C_3H_4$ adduct to products $C_3H_3 + H$ (Dixon, et al. 2001). As a result, we assume that this branching ratio does not change over the range 50 – 296 K. Secondly, we need to consider possible absorption of the VUV excitation and fluorescence intensities by residual $NH_3$ and $C_2H_4$ in the reactor. Although secondary absorption was found to be as high as 7 % at 296 K, no correction was required as both $NH_3$ and $C_2H_4$ were estimated to result in similar absorption losses. Below room temperature, the correction was less than 3% and it was always comparable for $NH_3$ and $C_2H_4$ within any single series of experiments. The temperature dependent H atom yields obtained in this way are listed in Table 2 leading to an approximately temperature independent value for the H atom yield of 0.99 ± 0.08. The yields at each temperature are the mean ratio of at least six pairs of *A* factors derived by fitting to H atom profiles similar to those shown in Figure 7.

Table 2 H atom yields for the $C(^3P)$ + $NH_3$ reaction as a function of temperature

| T / K | Number of experiments | H atom yield [a] |
|---|---|---|
| 296 | 12 | 0.99 ± 0.06 |
| 177 | 7 | 0.98 ± 0.08 |
| 106 | 8 | 0.92 ± 0.08 |
| 50 | 6 | 1.08 ± 0.14 |

[a]The error bars reflect the statistical uncertainties at the level of a single standard deviation including the uncertainties of the H branching ratio of the C + $C_2H_4$ reaction used as a reference.

A series of experiments was also conducted to examine the effect of pressure on the H atom yield of the C + $NH_3$ reaction at 296 K. In a similar manner to the temperature dependent H atom yield studies reported above, several pairs of H-atom curves were acquired at five different pressures over the range 1 - 20 Torr. The results (after consideration of the possible correction factors outlined above) are listed in Table 3.

Table 3 H atom yields at 296 K for the C($^3$P) + NH$_3$ reaction as a function of pressure

| P/ Torr | Number of experiments | H atom yield [a] |
|---|---|---|
| 1.030 | 4 | 1.05 ± 0.10 |
| 2.045 | 4 | 0.98 ± 0.10 |
| 5.039 | 4 | 0.98 ± 0.06 |
| 9.980 | 2 | 1.05 ± 0.14 |
| 19.66 | 4 | 1.03 ± 0.10 |

[a]The error bars reflect the statistical uncertainties at the level of a single standard deviation including the uncertainties of the H atom branching ratio of the C + C$_2$H$_4$ reaction used as a reference.

It is apparent from Table 3 that the absolute H atom yield does not vary as a function of pressure at 296 K, at least over the range of pressures investigated with a mean value for the H-atom yield of 1.02 ± 0.06. The absence of either pressure or temperature dependence for atomic hydrogen production leads us to the following conclusions on the basis of the electronic structure calculations. First, there does not seem to be any stabilization of the pre-reactive C⋯NH$_3$ complex in the van der Waals well of the various H$_3$CN adduct isomers. Any significant stabilization would result in lower H atom yields at both high pressure and at low temperature; neither of which are observed. Second, the determination of an H-atom yield close to unity at all pressures and temperatures investigated indicates that the C + NH$_3$ reaction leads mostly to H$_2$CN + H formation with only very small contributions from the HCN + H + H and HNC + H + H channels, which are predicted to be exothermic by -194 kJ/mol and -139 kJ/mol, respectively (using the enthalpy of formation values from (Baulch et al. 2005) for C, NH$_3$, H and HCN species and the (Hansel et al. 1998) value for HNC). Indeed, if the HCN/HNC formation channels were important, a value much closer to two would be obtained for the H atom yield. This result is somewhat surprising because the large exothermicity of the H$_2$CN + H channel (-300 kJ/mol considering the enthalpy of formation of H$_2$CN from (Jursic 1999; Zhou & Schlegel 2009)) could lead to secondary dissociation of the energized $H_2CN$. This result strongly suggests that the hydrogen atoms produced by $H_2CNH$ (or $HCNH_2$) dissociation carry an important fraction of the energy in the form of kinetic energy and then most, if not all the $H_2CN$ is formed below the barrier toward dissociation.

## 4. CONCLUSION

The product formation channels of ground state carbon atoms, $C(^3P)$, reacting with ammonia, $NH_3$, have been investigated combining complementary experiments and electronic structure calculations.  Electronic structure calculations of the intermediates, transition states and complexes along the reaction coordinate establish that the reaction is characterized by a submerged barrier. Product detection using time-resolved tunable VUV photoionization and time of flight mass spectrometry in a slow flow reactor ($T = 332$ K, $P = 4$ torr) shows the efficient formation of $H_2CN$ as a primary product of the reaction. H-atom yield measurements using the Laval nozzle technique from room temperature down to 50 K also indicate that hydrogen loss was the only exit channel of the $C(^3P) + NH_3$ reaction. The exothermic channels $HCN + H + H$ and $HNC + H + H$ are certainly minor products of the title reaction if they are produced at all. Furthermore, no stabilization of the pre-reactive $C \cdots NH_3$ complex in the van der Waals well of the various $H_3CN$ adduct isomers is observed experimentally down to 50 K. This work indicates therefore that in the temperature and pressure range investigated, the $H + H_2CN$ production channel represents 100% of the product yield for this reaction.

Despite the high abundance of both carbon atoms and ammonia in the interstellar medium, the $C(^3P) + NH_3$ reaction is currently absent from photochemical models. As demonstrated in part II of this work, the $C(^3P) + NH_3$ reaction is fast at room temperature and below, despite previous measurements to the contrary (Deeyamulla & Husain 2007). The title reaction is very likely to influence the chemistry of nitrogen hydrides, $NH_3$, $NH_2$ and NH, and more generally to affect nitrogen chemistry in the interstellar medium. The experimental temperature dependence determination of the rate coefficients as well as a study of their influence on interstellar nitrogen hydride abundances using a dense interstellar cloud model are reported in paper II.

**ACKNOWLEDGMENTS**

The Rennes team acknowledges support from the Agence Nationale de la Recherche, contract ANR-11-BS04-024-CRESUSOL-01, the French INSU/CNRS Program "Physique et Chimie du Milieu Interstellaire" (PCMI), the Institut National de Physique (INP CNRS), the Région Bretagne and the Université de Rennes 1. S.D.L.P. acknowledges financial support from the Institut Universitaire de France. F.G. and R.K.A.K acknowledge founding by the West Virginia University (startup package) for supply and travel support. Acknowledgement is also made to the Donors of the American Chemical Society Petroleum Research Fund for partial support of this research (PRF#53105-DN16 for R.K.A.K postdoctoral support). J.C.L. and K.M.H. acknowledge support from the French INSU/CNRS Programs PCMI and PNP. We thank Mr. Howard Johnsen and Dr. John Savee for technical support of this experiment. We also thank Dr. Doug Taube for his help and advice during the carbon suboxide synthesis. D.L.O. and the instrumentation for this work are supported by the Division of Chemical Sciences, Geosciences, and Biosciences, the Office of Basic Energy Sciences, the U. S. Department of Energy. Sandia is a multi-program laboratory operated by Sandia Corporation, a Lockheed Martin Company, for the National Nuclear Security Administration under contract DE-AC04-94-AL85000. VW research is supported by the ERC Starting Grant 3DICE (grant agreement 336474) and the the French INSU/CNRS Program "Physique et Chimie du Milieu Interstellaire" (PCMI). This research used resources of the Advanced Light Source, a DOE Office of Science User Facility, which is supported by the Direct, Office of Science, Office of Basic Energy Sciences, the U.S. Department of Energy under contract DE-AC02-05CH11231 at Lawrence Berkeley National Laboratory.

# Appendix A: Chemical modeling of the flow experiments.

Table 4 shows all the reactions included in the chemical modeling of the flow experiments. Some of the reactions treated in this table have unknown reaction rates. For radical-radical reactions we assume that reactions proceeding on PESs arising from pairing of the two radical unpaired electrons do not show any barrier, whereas if these electrons remain unpaired the surface is likely to be repulsive. For example reactions between two doublet radicals are considered to have no barrier for the singlet surface but a barrier for the triplet surface. For carbon atom reactivity with closed shell molecules, the estimates are based on the widely available experimental and theoretical studies. Rate constants are estimated using capture rate theory dominated by long-range forces, mainly through dispersion interactions (Clary et al. 1994; Harding, Georgievskii, & Klippenstein 2005) taking into account the electronic degeneracy.

**Table 4** Reactions and rate constants used in the chemical simulations (reactions in grey are found to play a minor role under the experimental conditions used in this study):

| | Reaction | | k(300K) | ref |
|---|---|---|---|---|
| 1. | H + CH | → C + $H_2$ | 1.24e-10 | KIDA datasheet, http://kida.obs.u-bordeaux1.fr/ |
| 2. | H + $H_2$CN | → HCN + $H_2$ | 2.0e-10 | (Hébrard et al. 2012; Nesbitt, Marston, & Stief 1990) |
| 3. | H + $^{1,3}C_2O$ | → CH + CO | 3.0e-11 | (Bauer, Becker, & Meuser 1985; Horie et al. 1983; Schmidt et al. 1983) |
| 4. | C + $NH_2$ | → HCN + H | 6.0e-11 | (Herbst, Terzieva, & Talbi 2000; Talbi 1999) |
| 5. | C + $NH_3$ | → $H_2$CN + H | 8.0e-11 | Paper II |
| 6. | C + $^{1,3}C_2O$ | → $C_2$ + CO | 4.0e-11 | By comparison with C + $O_2$ (triplet + triplet reaction) |
| 7. | C + $C_3O_2$ | → $C_2$ + CO + CO | 2.0e-10 | (Deeyamulla & Husain 2006) |
| 8. | C + $CH_2$NH | → HCN + $CH_2$ | 2.0e-10 | Estimated from (Feng et al. 2007) and (Harding, et al. 2005) |
| 9. | C + $H_2$CN | → HCN + CH | 3.00e-11 | Estimated from (Cho & Andrews 2011; Koput 2003; Lau et al. 1999; Osamura & Petrie 2004) |
| | | → H + HCCN | 3.00e-11 | |
| 10. | CH + $NH_3$ | → $CH_2$NH + H | 1.69e-10 | KIDA datasheet (Blitz et al. 2012; Bocherel et al. 1996; Zabarnick, Fleming, & Lin 1989) |
| 11. | CH + $NH_2$ | → $HCNH_2$ | 0 | |
| | | → HCN + H + H | 2.0e-10 | |
| 12. | CH + $C_3O_2$ | → $C_2$H + CO + CO | 1.0e-10 | (Sato et al. 1998) propose $1.0^e$-11 deduced from a complex mechanism. |
| 13. | CH + $^{1,3}C_2O$ | → $C_2$H + CO | 2.0e-10 | |
| 14. | $C_2$ + $NH_3$ | → $C_2$H + $NH_2$ | 3.9e-11 | / $C_2$H + $NH_3$ and $C_2$ reactions (Canosa et al. 2007; |

| | | | |
|---|---|---|---|
| | | | Daugey, et al. 2008; Paramo et al. 2008; Reisler, Mangir, & Wittig 1980) |
| 15. $C_2 + NH_2$ | → products | | Minor reaction leading very likely not to $H_2CN$ |
| 16. $C_2 + C_3O_2$ | → $C_3 + CO + CO$ | 1.6e-10 | (Becker et al. 2000) |
| 17. $C_2H + NH_3$ | → $C_2H_2 + NH_2$ | 3.9e-11 | (Carl et al. 2004) |
| 18. $C_2H + NH_2$ | → products | | Minor reaction leading very likely not to $H_2CN$ |
| 19. $C_2H + C_3O_2$ | → $HC_4O + CO$ | 2.0e-11 | Estimated by comparison with $C_2H$ + alkenes, alkynes reactions. |
| 20. $C_2H + {}^{1,3}C_2O$ | → $C_3H + CO$ | 4.0e-11 | Estimated by comparison with $C_2H + NO, O_2$ reactions. |
| 21. $H_2CN + NH_2$ | → $HCN + NH_3$ | 5.4e-11 | (Yelle et al. 2010) |
| 22. ${}^{1,3}C_2O + NH_2$ | → $H_2CN + CO$ <br> → $H + HCN + CO$ | 1.0e-11 <br> 3.0e-11 | Considering no barrier (radical + radical reaction). |
| 23. ${}^1C_2O + NH_3$ | → $CH_2NH + CO$ <br> → $HCN + H_2 + CO$ | 3.0e-11 <br> 1.6e-11 | |
| 24. ${}^1C_2O + C_3O_2$ | → $C_2 + CO + CO + CO$ | 1.0e-10 | By comparison with ${}^1C_2O + NH_3$ |
| 25. $NH_2 + NH_2$ | → $N_2H_2 + H_2$ <br> → $N_2H_4$ | 0 <br> $k_0$=5.7e-29 <br> $k_\infty$=8.0e-11 | (Altinay & Macdonald 2012; Asatryan et al. 2010; Bahng & Macdonald 2008; Fagerstrom et al. 1995; Klippenstein et al. 2009; Stothard, Humpfer, & Grotheer 1995) |
| 26. $H_2CN + H_2CN$ | → $HCN + H_2CNH$ | 7.7e-12 | (Nizamov & Dagdigian 2003) |

**Appendix B: Ionization cross sections**

The photoionization cross sections of the different species were estimated using the model of Koizumi (Koizumi 1991; Rio et al. 2010). All details can be found in (Koizumi 1991; Rio, et al. 2010). We also compare these photoionization cross sections to those inferred from the model of Bobeldijk, Zande, and Kistemaker (1994) based on the additivity concept. Table 5 presents experimental adiabatic Ionization Energies ($IE_{ad,exp}$) and vertical Ionization Energies ($IE_{vert}$, calculated at the EPT/cc-pVTZ level of calculation for ground and excited ionic states). The number in brackets in the $IE_{th,vert}$ column is the number of electrons involved in each ionization step, $\sigma_{model}$ is the value used in the simulation (ionization at 13.7 eV). The cross section $\sigma_{Koizumi}$ is the value given by the Koizumi model and $\sigma_{Bobeldijk}$ is the value given by the Bobeldijk model. Photoionization cross section are in Mb=$10^{-18}$ cm$^2$.

**Table 5** Comparison of photoionization cross sections obtained by different models.

| Species | M (g) | IE$_{ad,exp}$ (eV) | IE$_{th,vert}$ (eV) Ionic state1/state2/state3/… | $\sigma_{Koizumi}$ @13.7 eV | $\sigma_{Bobeldijk}$ @11.8 eV | $\sigma_{Bobeldijk}$ @16.7 eV | References |
|---|---|---|---|---|---|---|---|
| NH$_3$ | 17 | 10.07 | 10.75(2), 16.48(2) | 8.4 | 9 | 27 | (Edvardsson et al. 1999) |
| NH$_2$ | 16 | 11.17 | 11.89(1), 12.48(1), 12.97(1), 16.56(1) | 16 | 6 | 18 | (Dixon, et al. 2001) |
| HCN | 27 | 13.60 | 13.68 (2), 14.03 (2) | 12 | | | (Dibeler, Reese, & Franklin 1961; Nuth & Glicker 1982) |
| C$_2$H$_2$ | 26 | 11.400 | 11.32(4), 17.26 | 18 | 25 | 25 | (Cool et al. 2005) |
| H$_2$CN | 28 | | 10.87(1), 12.54(1), 13.01(1), 13.10(1), 14.31(1) | 22 | 2 | 26 | (Nesbitt et al. 1991b) |
| CH$_2$NH | 29 | 10.0 | 10.70(2), 12.42(2), 15.18(2) | 19 | 5 | 35 | (Peel & Willett 1975) |
| $^3$C$_2$O | 40 | | 11.14(1), 11.26(1), 12.80(1),16.34(1) | 15 | 21 | 23 | |
| $^1$C$_2$O | 40 | | 10.35(2), 12.38(2), 16.85(2) | 19 | 21 | 23 | |

**Appendix C: Methylenimine photoionization**

Methylenimine (CH$_2$NH) is an important intermediate in the C + NH$_3$ system. Its adiabatic and first three vertical ionization energies have been determined using He I photoelectron spectrum (Peel & Willett 1975). At 13.7 eV the two first IEs are accessible. Considering the very high stability of HCNH$^+$ ions, we consider the possibility of the CH$_2$NH$^+$ → HCNH$^+$ + H dissociation. At the M06-2X/cc-pVTZ level the HCNH$^+$ + H dissociation is found to occur on the CH$_2$NH$^+$ ground state through a transition state located at 11.27 eV above the neutral CH$_2$NH ground state as shown in Figure 11. We consider that the first excited state of CH$_2$NH$^+$, located above the dissociation transition state, leads to HCNH$^+$. Considering the respective area of the peak in He I photoelectron spectrum, almost 40% of the ionization of CH$_2$NH at 13.7 eV should lead to m/z = 28. Consequently, most of the m/z = 28 ion peak observed in the flow experiments after 3 ms comes from CH$_2$NH$^+$, the direct H$_2$CN photoionization contribution being confined to short times. It should be noted that the calculated adiabatic value of the IE (10.00 eV) as well as the calculated first three vertical IEs (10.70, 12.42 and 15.18 eV) are in good agreement with the observed ones (10.0 eV for the adiabatic IE, 10.52, 12.43, and 15.13 for the vertical ones).

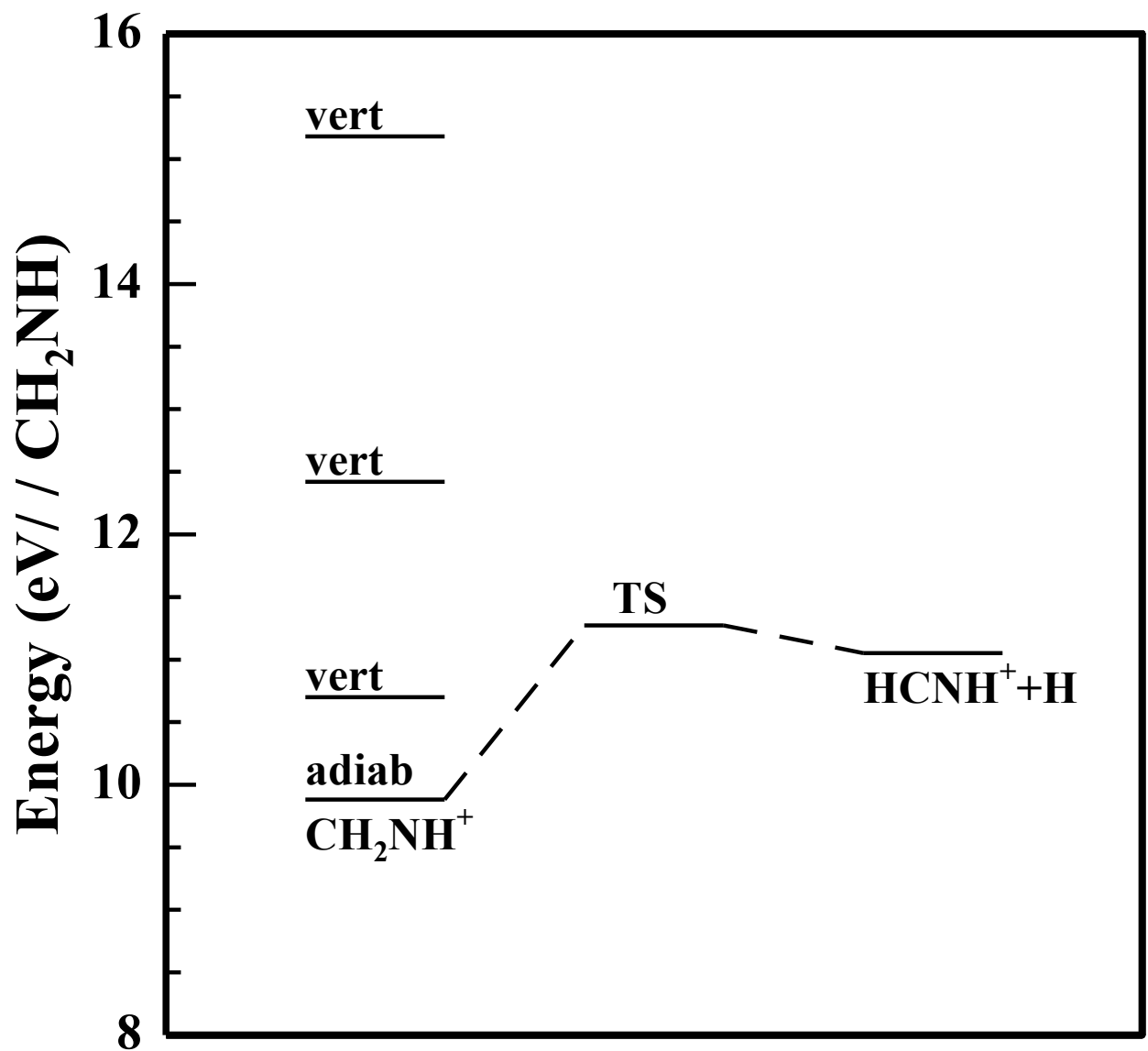

**Figure 8.** Potential energy diagram for the $CH_2NH^+$ dissociation calculated at the M06-2X/cc-pVQZ level.

# The C($^3$P) + NH$_3$ reaction in interstellar chemistry:
# I. Investigation of the product formation channels
*Bourgalais et al.*

## Online-only supplementary materials: C + NH$_3$ theoretical calculations

**Table S1** Reaction enthalpies (at 0 K including ZPE) of species involved in the C + NH$_3$ reaction calculated by different methods (zero energy is the C + NH$_3$ level).

|  | CNH$_3$ | TS1 | TS2 |
|---|---|---|---|
| MRCI+Q/avtz//CASCCF/avtz[a] | -101 | -13 | -7 |
| MRCI+Q/avqz//CASCCF/avqz[b] | -103 | -16 | -10 |
| CCSDT/avtz | -95 | -6 | +5 |
| CCSDT/avqz | -98 | -10 | 0 |
| M06/vqz | -119 | -38 | -31 |

(a) ZPE at CCSDT/avtz level
(b) ZPE at CCSDT/avqz level

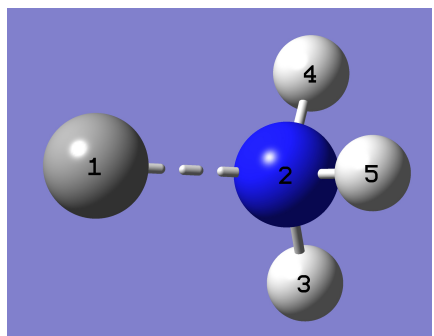

**Figure S1** Initial C and NH$_3$ reaction intermediate.

**Table S2** Distances (Å), angles (degrees), and Frequencies (cm$^{-1}$) for the CNH$_3$ reaction adduct

|  |  | M06-2X/cc-pVQZ | CCSD(T)/aug-cc-pVQZ |
|---|---|---|---|
| Bond length (Å) | C$_1$-N$_2$ | 1.6036 | 1.6611 |
|  | H$_3$-N$_2$ | 1.0180 | 1.0257 |
|  | H$_4$-N$_2$ | 1.0180 | 1.0257 |
|  | H$_5$-N$_2$ | 1.0180 | 1.0257 |
|  | H$_3$-N$_2$-C$_1$ | 110.84 | 110.74 |
|  | H$_4$-N$_2$-C$_1$ | 110.84 | 110.74 |
|  | H$_5$-N$_2$-C$_1$ | 110.84 | 110.74 |
| Angle (degrees) | Diedre(H$_4$-N$_2$-C$_1$-H$_3$) | 120.00 | 120.00 |
|  | Diedre(H$_5$-N$_2$-C$_1$-H$_3$) | -120.00 | -120.00 |
| Frequencies (cm$^{-1}$) |  | 593, 774, 774, 1307, 1635, 1635, 3400, 3502, 3502 | 564, 776, 776, 1311, 1653, 1653, 3409, 3523, 3523 |

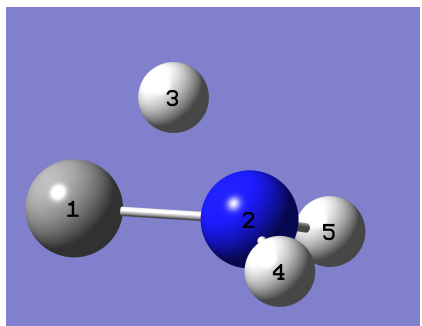

**Figure S2** Geometry of TS1

**Table S3** Distances (Å), angles (degrees), and Frequencies (cm$^{-1}$) for TS1

| | | M06-2X/cc-pVQZ | CCSD(T)/aug-cc-pVQZ |
|---|---|---|---|
| Bond length (Å) | $C_1$-$N_2$ | 1.5359 | 1.5768 |
| | $H_3$-$N_2$ | 1.2169 | 1.2415 |
| | $H_4$-$N_2$ | 1.0139 | 1.0226 |
| | $H_5$-$N_2$ | 1.0139 | 1.0226 |
| | $H_3$-$N_2$-$C_1$ | 56.21 | 55.0 |
| | $H_4$-$N_2$-$C_1$ | 118.70 | 118.80 |
| | $H_5$-$N_2$-$C_1$ | 118.70 | 118.80 |
| Angle (degrees) | Diedre($H_4$-$N_2$-$C_1$-$H_3$) | 106.07 | 105.40 |
| | Diedre($H_5$-$N_2$-$C_1$-$H_3$) | -106.07 | -105.40 |
| Frequencies (cm$^{-1}$) | | 1608i, 682, 838, 892, 1110, 1510, 2226, 3447, 3579 | 1691i, 672, 836, 841, 1136, 1520, 2219, 3461, 3596 |

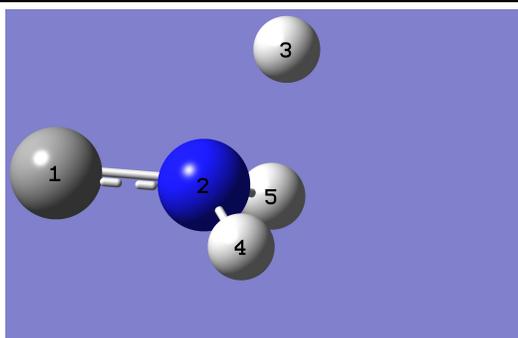

**Figure S3** Geometry of TS2

**Table S4** Distances (Å), angles (degrees), and Frequencies (cm$^{-1}$) for TS2

|  |  | M06-2X/cc-pVQZ | CCSD(T)/aug-cc-pVQZ |
|---|---|---|---|
| Bond length (Å) | $C_1$-$N_2$ | 1.3445 | 1.3752 |
|  | $H_3$-$N_2$ | 1.5002 | 1.4920 |
|  | $H_4$-$N_2$ | 1.0209 | 1.0287 |
|  | $H_5$-$N_2$ | 1.0209 | 1.0287 |
|  | $H_3$-$N_2$-$C_1$ | 117.98 | 118.40 |
|  | $H_4$-$N_2$-$C_1$ | 118.60 | 118.00 |
|  | $H_5$-$N_2$-$C_1$ | 118.60 | 118.00 |
| Angle (degrees) | Diedre($H_4$-$N_2$-$C_1$-$H_3$) | 109.50 | 110.40 |
|  | Diedre($H_5$-$N_2$-$C_1$-$H_3$) | - 109.50 | -110.40 |
| Frequencies (cm$^{-1}$) |  | 1431i, 705, 910, 1092, 1108, 1266, 1592, 3346, 3408 | 1783i, 708, 911, 1091, 1110, 1240, 1601, 3369, 3435 |